\documentclass[twocolumn,preprintnumbers,amsmath,amssymb,superscriptaddress,prb]{revtex4}

\usepackage{graphicx,amsmath,amsfonts,amssymb}
\usepackage{graphics}
\usepackage{subfig}
\usepackage{dcolumn}
\usepackage{bm}
\usepackage{color}

\begin{document}

\title{Intrinsic Energy Dissipation in CVD-Grown Graphene Nanoresonators}

\author{Zenan Qi}
    \affiliation{Department of Mechanical Engineering, Boston University, Boston, MA 02215}
\author{Harold S. Park\footnote{Electronic address: parkhs@bu.edu}}
    \affiliation{Department of Mechanical Engineering, Boston University, Boston, MA 02215}

\date{\today}

\begin{abstract}

We utilize classical molecular dynamics to study the the quality (Q)-factors of monolayer CVD-grown graphene nanoresonators.  In particular, we focus on the effects of intrinsic grain boundaries  of different orientations, which result from the CVD growth process, on the Q-factors.  For a range of misorientations orientation angles that are consistent with those seen experimentally in CVD-grown graphene, i.e. 0$^{\circ}$ to $\sim$20$^{\circ}$, we find that the Q-factors for graphene with intrinsic grain boundaries are 1-2 orders of magnitude smaller than that of pristine monolayer graphene.  We find that the Q-factor degradation is strongly influenced by both the symmetry and structure of the 5-7 defect pairs that occur at the grain boundary.  Because of this, we also demonstrate that find the Q-factors CVD-grown graphene can be significantly elevated, and approach that of pristine graphene, through application of modest (1\%) tensile strain.

\end{abstract}

\maketitle

\section{Introduction}

Since its recent discovery as the simplest two-dimensional crystal structure~\cite{novoselovNATURE2005}, graphene has been extensively studied not only for its unusual physical properties resulting from its two-dimensional structure~\cite{geimNM2007,netoRMP2009,hanPRL2007,seolSCIENCE2010}, but also for its potential as the basic building block of future applications, i.e. nanoelectromechanical systems (NEMS)~\cite{bunchSCIENCE2007,bunchNL2008,robinsonNL2008a,bartonJVSTB2011,eomPR2011}.

Graphene is viewed as an ideal material for NEMS-based sensing and detection applications due to its combination of extremely low mass and exceptional mechanical properties~\cite{leeSCIENCE2008}; we note the recent review of Barton \emph{et al.}~\cite{bartonJVSTB2011} in this regard.  However, one key issue limiting the applicability of graphene as a sensing component is its low quality (Q)-factor; the Q-factors of a 20-nm thick multilayer graphene sheet were found to range from 100 to 1800 as the temperature decreased from 300 K to 50 K~\cite{bunchSCIENCE2007}.  Similarly low Q-factors between 2 and 30 were also observed by Sanchez \emph{et al.}~\cite{sanchezNL2008} for multilayer graphene sheets, while higher Q-factors with values up to 4000 were reported using multilayer graphene oxide films~\cite{robinsonNL2008a}.  Theoretically, the Q-factors of graphene were recently studied using classical molecular dynamics (MD) simulations~\cite{kimNL2009,kimAPL2009}, where spurious edge modes that are present in suspended graphene were proposed to have a key role in the low Q-factors that were observed experimentally.  This hypothesis was recently validated by Barton \emph{et al.}~\cite{bartonNL2011}, who found Q-factors approaching 2000 for graphene resonators that were clamped on all slides, thus eliminating the spurious edge modes.  

These early experimental works on graphene nanoresonators utilized graphene flakes made via the scotch-tape method, which produces graphene sheets of varying thickness, though importantly, each graphene layer in the sheet is single crystalline.   However, graphene research has been transformed by the recent development of the chemical-vapor-deposition (CVD) growth process to synthesize large area graphene sheets~\cite{liSCIENCE2009,parkNN2009,reinaNL2009}.  CVD technology has the potential to revolutionize graphene-based sensing technology due to the resulting promise of wafer-scale graphene devices comprised of large arrays of single-layer graphene resonators~\cite{zandeNL2010}.  However, while large area graphene films are desirable for these and other graphene-based applications~\cite{kimNATURE2009}, the graphene films that result from CVD growth are polycrystalline, and thus are composed of many interconnected single crystalline graphene grains that intersect at grain boundaries having a range of misorientation angles~\cite{huangNATURE2011}, which generally are less than 20$^{\circ}$~\cite{Nemes-Incze2011}.  While these grain boundaries are sometimes viewed favorably for tunable electronic devices~\cite{Carr2010}, they are likely to have a deleterious effect for graphene nanoresonators because the misorientation and the resulting non-ideal bonding at the grain boundary causes an increase in phonon scattering, which creates another energy dissipation mechanism and a lower Q-factor.

Some very recent experimental studies, such as those of van der Zande \emph{et al.}~\cite{zandeNL2010} and Barton \emph{et al.}~\cite{bartonNL2011} have studied CVD-grown graphene nanoresonators, and have found relatively high Q-factors on the order of about 2000.  However, those works also improved the Q-factors by removing spurious edge modes~\cite{kimNL2009}, and therefore the intrinsic losses that occur in CVD-grown graphene due to the existence of the grain boundaries is unknown.  Furthermore, previous theoretical studies~\cite{kimNL2009,kimAPL2009,kimNANO2010} focused on the energy dissipation mechanisms in pure graphene without grain boundaries.  Therefore, the objective of the present work is to quantify, via classical MD simulations, the intrinsic dissipation mechanisms introduced in CVD-grown graphene nanoresonators due to the presence of the grain boundaries.  

\section{Simulation Methodology} 

\begin{figure} \begin{center} 
\includegraphics[scale=0.35]{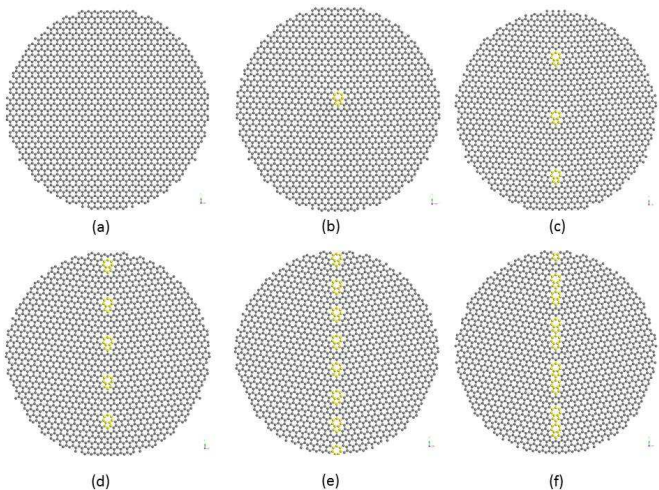}
\caption{\label{}(Color online)  Schematic of the graphene monolayers containing grain boundaries with various misorientation angles of (a) 0$^{\circ}$;(b)1.4$^{\circ}$; (c) 5.31$^{\circ}$; (d) 9.83$^{\circ}$;  (e) 12.83$^{\circ}$;  (f) 16.62$^{\circ}$.  5-7 defect pairs are highlighted in yellow.}
\label{schematic} \end{center} \end{figure}

The starting point of our simulations is to note that recent experimental studies have found that the misorientation angles at grain boundaries in CVD-grown graphene lie mostly between 0$^{\circ}$ to 20$^{\circ}$~\cite{Nemes-Incze2011}.  To systematically study the effects of different grain boundary orientations on the intrinsic loss mechanisms in graphene monolayers, we created graphene monolayers with a single grain boundary along the center line of the monolayer that runs along the armchair orientation with 6 different misorientation angles of 0$^{\circ}$ (pristine graphene), 1.4$^{\circ}$, 5.31$^{\circ}$, 9.83$^{\circ}$, 12.83$^{\circ}$ and 16.62$^{\circ}$.  These configurations are shown in Fig. \ref{schematic}, where the diameter of all the graphene monolayers was 4 nm.  The initial configuration including the single grain boundary was generated by rotating two semicircular graphene monolayers and then piecing them together.  At that point, energy minimization using the conjugate gradient algorithm was employed to optimize the structure, where the carbon-carbon interactions were modeled using the AIREBO potential of~\citet{stuartJCP2000}, which is able to accurately simulate the forming and breaking of carbon bonds.  A key point to note is that the creation of the grain boundary results in the formation of 5-7 defect ring pairs, which are colored in yellow in Fig. \ref{schematic}.  We can observe that as the grain boundary misorientation angle increases, the density of the 5-7 ring pairs increases.  For example, for the smallest misorientation angle of 1.4$^{\circ}$ in Fig. \ref{schematic}(a), there exists only a single 5-7 unit ring pair near the center of the monolayer.  In contrast, as the grain boundary misorientation angle increases, the density of the 5-7 ring pairs increases, as observed for the other cases in Fig. \ref{schematic}.

The Q-factors were calculated and the intrinsic energy dissipation mechanisms studied using classical MD via the publicly available simulation code LAMMPS~\cite{plimptonLAMMPS}.  After obtaining the equilibrium graphene monolayer structures with the various grain boundary misorientation angles as shown in Fig. \ref{schematic}, we then performed a thermal equilibration using a Nose-Hoover thermostat~\cite{hooverPRA1985} for 500 ps using a time step of 1 fs, i.e. within an NVT ensemble. During the equilibration, the edges of the graphene sheet were constrained in plane while the rest of the sheet was left free to move.  Previous theoretical~\cite{kimNL2009} and experimental~\cite{bartonNL2011} works have demonstrated that spurious edge vibrational modes, which arise due to the undercoordinated nature of bonding at the edges of graphene, have a dominant role in reducing the Q-factors of suspended graphene nanoresonators; these edge modes would also be present for suspended CVD-grown nanoresonators, and thus should be eliminated to maximize the Q-factor.  Because of this, after the thermal equilibration, the edges of the graphene monolayer were clamped at the equilibrium diameter that is established during the thermal equilibration to eliminate the possibility of spurious edge modes. 

\begin{figure} \begin{center} 
\includegraphics[scale=0.25]{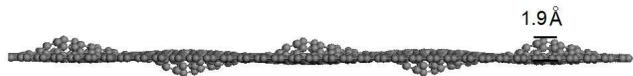}
\caption{\label{}Out-of-plane buckling of 9.83$^{\circ}$ defective graphene sheet after thermal equilibration.  Visualization performed using VMD~\cite{Humphrey1996}.}
\label{buckle} \end{center} \end{figure}

During the thermal equilibration, out-of-plane buckling was observed due to the 5-7 defect pairs as observed in Fig. \ref{buckle}; such buckles were previously observed in MD simulations by~\citet{Liu2011}.  We find that for the smallest grain boundary angle of 1.4$^{\circ}$ at 3K, the height of the buckle is about 3.3\AA. For larger grain boundary misorientation angles, there are more buckles along the grain boundary due to the larger number of 5-7 defect pairs, which interact and lead to a decrease in the buckling height.  The impact of the buckles on the Q-factors will be elucidated later; in particular, we will demonstrate the utility of tensile mechanical strain in enhancing the Q-factor by flattening out the buckles.

After the thermal equilibration, the graphene monolayer was actuated by assigning an initial sinusoidal velocity profile that ranged from zero at the clamped edges to a maximum at the center of the circular monolayer, and where the initial velocity was applied only in the vertical $z$-direction to be perpendicular to the graphene sheets, and was chosen to be sufficiently small such that the resulting oscillation of the graphene monolayer would be purely harmonic, i.e. the resulting increase in total energy due to the applied sinusoidal velocity was only about 0.1\%. While the buckling results in a non-planar graphene monolayer, for consistency, the direction of the applied initial velocities were the same for both pristine and defective graphene sheets. After the velocity profile was prescribed, the resulting free oscillation of the graphene monolayer was performed within an energy conserving (NVE) ensemble for 3000 ps.  

\begin{figure} \begin{center} 
\includegraphics[scale=0.3]{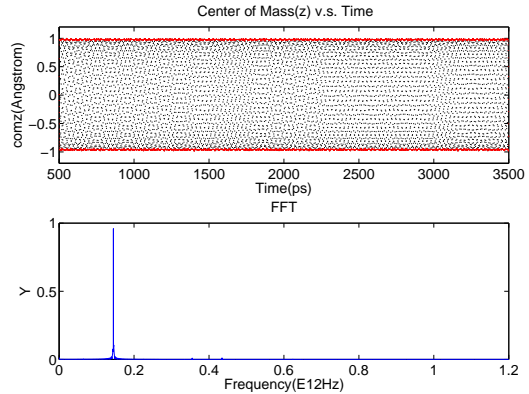}
\caption{\label{}Center of mass and natural frequency for pristine graphene at 3K.}
\label{comp} \end{center} \end{figure}

\begin{figure} \begin{center} 
\includegraphics[scale=0.3]{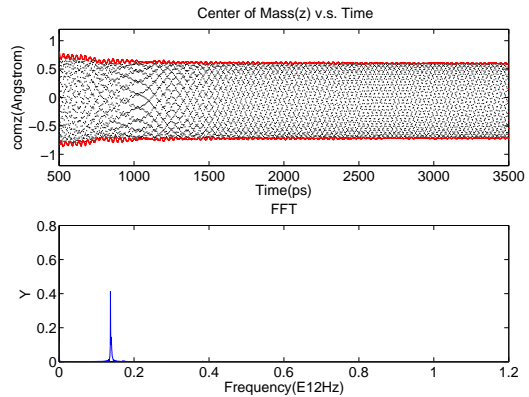}
\caption{\label{}Center of mass and natural frequency for defective graphene with a grain boundary misorientation angle of 5.31$^{\circ}$ at 3K.}
\label{comd} \end{center} \end{figure}

The Q-factors were calculated following the procedure described by~\citet{valJAP2011} and~\citet{chuARXIV2007}.  Specifically, as described by~\citet{valJAP2011}, the variation of the displacement of the center of mass of the graphene monolayer was tracked for the duration of the simulation after the initial velocity is applied.  The decay in the root mean square displacement was then fit to the following exponential curve ($e^{-\gamma\omega t}$), which is then related to Q via Q$=0.5/\zeta$, where $\zeta=\gamma/\omega$ is the damping ratio and $\omega$ is the angular vibrational frequency.  Specifically, because the vibrational motion is predominately in the $z$-direction in our MD simulations, with little contribution from the motion in the $x$ and $y$ directions, we used the center of mass in the $z$-direction only to fit the damping curve. This approach is utilized in the present work as it avoids the necessity of extracting the external energy as previously performed by~\citet{kimNL2009,kimAPL2009,kimNANO2010}.  Figs. \ref{comp} and \ref{comd} show typical center of mass and natural frequency results for pristine and 5.31$^{\circ}$ defective graphene sheets, respectively, that were used to obtain the Q-factors.

We make three other relevant comments here.  First, we chose to study graphene monolayers with a single grain boundary rather than study polycrystalline graphene with a distribution of grain boundary misorientation angles.  By comparing the Q-factors of monolayer graphene with a single grain boundary and by varying the misorientation angle of the single grain boundary to pristine graphene, and by utilizing temperatures ranging from $\sim$0K to 300K, we aim to quantify the effects of each grain boundary orientation on the intrinsic loss mechanisms, or Q-factor.  Second, the models in this work represent extreme cases in that the defects run through the entire diameter of the graphene monolayer.  Finally, we will also study the effects of tensile mechanical strain in enhancing the Q-factors of graphene.  Previous studies on both graphene~\cite{kimNL2009,kimNANO2010} and other nanostructures~\cite{verbridgeNL2007} have demonstrated the effectiveness of strain in increasing the Q-factors.  We will demonstrate the utility of strain in mitigating the effects of out of plane buckling due to the 5-7 unit ring defects along the grain boundaries, thus elevating the Q-factors.

\section{Numerical Results and Discussion}

Before presenting the misorientation angle-dependent results for the Q-factors, we first note again that the 5-7 defects along the grain boundary induce out of plane buckling in the graphene monolayer as shown in Fig. \ref{buckle}.  More specifically, we first show that both the orientation of the buckle (i.e. up or down), and its location along the grain boundary, significantly impacts not only the structural symmetry of the graphene sheet, but also the Q-factors; this is true for all grain boundary misorientation angles.

\begin{figure} \begin{center} 
\includegraphics[scale=0.36]{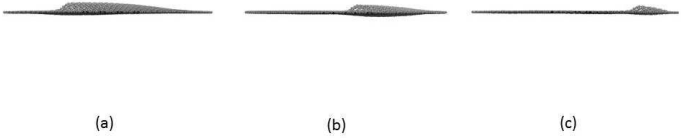}
\caption{\label{}(Color online)  Schematic figures showing out-of-plane bucklings of 1.4$^{\circ}$ sheet with different defect pair positions after relaxation.}
\label{bucknostrainall} \end{center} \end{figure}

To demonstrate this, we considered several possible spatial locations of the single 5-7 defect pair for the 1.4$^{\circ}$ misorientation angle, as illustrated in Fig. \ref{bucknostrainall}.  The first two cases, i.e. those depicted in Figs. \ref{bucknostrainall}(a) and (b), result in a significant asymmetry of the actuated oscillation.  However, the buckle in the third case in Fig. \ref{bucknostrainall}(c) is near the edge of the graphene monolayer, and thus better preserves the circular oscillation symmetry, and results in a Q-factor of about 10,000, which is almost 1 order of magnitude larger than found for the cases in Fig. \ref{bucknostrainall}(a) and (b).  However, to make the results of all the grain boundary misorientation cases to be comparable, we have placed the single 5-7 defect pair for the 1.4$^{\circ}$ case in the center of the monolayer as shown in Fig. \ref{schematic}(a).

\begin{figure} \begin{center} 
\includegraphics[scale=0.36]{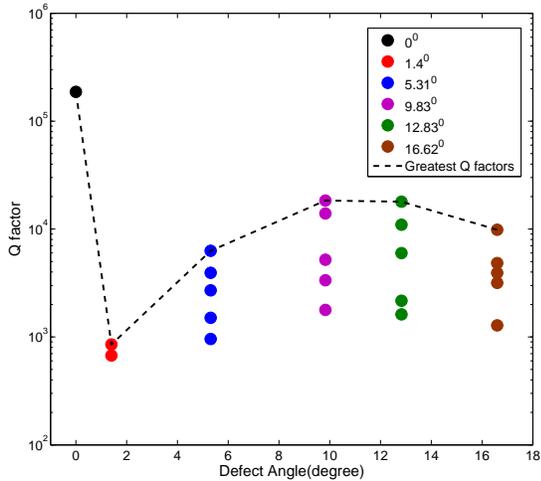}
\caption{\label{}(Color online)  Variation of the Q-factor for monolayer graphene at 3K for different grain boundary misorientation angles.}
\label{nostrainq3K} \end{center} \end{figure}

To illustrate the influence of the orientation of the buckling, for each misorientation angle (except for the 1.4$^{\circ}$ sheet with one buckle which thus has only two possible buckling orientations, up or down), we tested five cases with different buckling patterns at 3K, where the different buckling patterns were generated by using different random velocity seeds during the thermal equilibration.  From the results shown in Fig. \ref{nostrainq3K}, we can see that all cases have a Q-factor that is 1-2 orders of magnitude smaller than pristine, monolayer graphene (0$^{\circ}$).  Furthermore, the range between the highest and lowest Q-factor for each misorientation angle spans approximately one order of magnitude.  For consistency, we utilize the configuration with the highest Q-factor for each misorientation angle, as indicated by the dashed line in Fig. \ref{nostrainq3K}, for the remainder of this study.

\begin{figure} \begin{center} 
\includegraphics[scale=0.36]{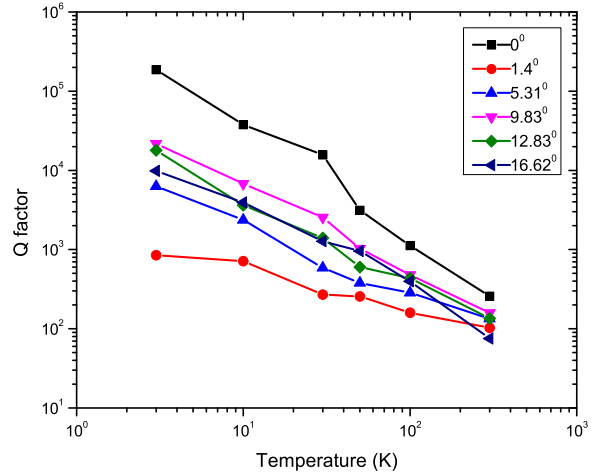}
\caption{\label{}(Color online) Variation of the Q-factor for unstrained monolayer graphene as a function of temperature and grain boundary misorientation angle.}
\label{nostrainq} \end{center} \end{figure}

The Q-factors of unstrained graphene, both pristine and with the different grain boundary orientations, are shown in Fig. \ref{nostrainq} as a function of temperature.  We find that the Q-factors of pristine graphene follow the relationship $Q\sim T^{-\alpha}$, where we find $\alpha=1.28$ for the present case, which is similar to the result of~\citet{kimNL2009} when the different approaches for calculating Q are accounted for.  We can see that for all temperatures, pristine graphene has a higher Q-factor than CVD graphene, where the difference in Q-factor at low temperatures is between 1-2 orders of magnitude, as shown in Fig.  \ref{nostrainq3K}.  We also observe that the Q factors for all graphene sheets with grain boundaries obey a similar functional form ($Q\sim T^{-\alpha}$) with respect to temperature, while for both pristine and defected graphene sheets the Q factor drops rapidly by the time room temperature is reached.

More interestingly, we find that the relationship between the grain boundary misorientation angle and the Q-factor is non-monotonic.  This is illustrated in Fig. \ref{nostrainq3K}, where the Q-factor is seen to first increase with increasing misorientation angle and then decrease when the angle becomes larger than about 10$^{\circ}$.  We believe this is due to a competition between two effects.

First, as the grain boundary misorientation angle increases, the density of 5-7 defect pairs along the grain boundary increases, which should result in a decrease in the Q-factor with increasing grain boundary misorientation angle.  On the other hand, while there are more buckles in the graphene monolayer with increasing grain boundary misorientation angle, we find that due to the interaction of the increasing number of defect pairs with increasing grain boundary misorientation angle that the circular symmetry of the graphene monolayer is better preserved.  Specifically, as seen in Fig. \ref{schematic} for the 1.4$^{\circ}$ case, a single, non-symmetric 5-7 defect pair exists near the center of the graphene sheet, which we have already discussed results in a substantial reduction in Q-factor.  However, increasing the misorientation angle results in more 5-7 defects on each side of the central 5-7 defect, which enhances the overall symmetry of the graphene sheet.

\begin{figure} \begin{center} 
\includegraphics[scale=0.14]{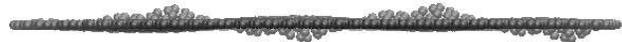}
\caption{\label{}(Color online) Out-of-plane buckling of 16.62$^{\circ}$ defective graphene sheet after thermal equilibration.}
\label{buckle16} \end{center} \end{figure}

Furthermore, the height of the buckles tends to decrease with increasing grain boundary misorientation angle, i.e. the buckling heights for the 1.4$^{\circ}$, 5.31$^{\circ}$, 9.83$^{\circ}$, 12.83$^{\circ}$ and 16.62$^{\circ}$ graphene sheets at 3K after relaxation are found to be 3.3\AA, 2.5\AA, 1.9\AA, 1.5\AA and 2.2\AA, respectively. Thus, the increase in Q-factor due to the greater structural symmetry with increasing misorientation angle coupled with the corresponding reduction in buckling height counteracts the decrease in Q due to the increase in the defect density. To illustrate this concept, we show in Fig. \ref{buckle16} the equilibrium buckled configuration for the 16.62$^{\circ}$ case, where it can be seen that each of the four buckles forms from the combination of two smaller buckles.  As a result, the buckles not only have larger buckling heights, but also break symmetry due to the fact that each buckle is composed of a smaller and larger sub-buckle.  These factors couple to result in a smaller Q-factor for the 16.62$^{\circ}$ case, and the overall non-monotonic trend seen in Fig. \ref{nostrainq}.

Due to the deleterious effect of the grain boundaries on the Q-factor, we also examine how the Q-factors of CVD-grown graphene can be enhanced.  As suggested in various works~\cite{zandeNL2010,kimNL2009,kimNANO2010,eomPR2011}, tensile mechanical strain is an effective approach to enhancing the Q-factors of nanostructures and NEMS.  In our MD simulations, we imposed a modest, experimentally-accessible 1\% tensile strain~\cite{Huang2009,Mohiuddin2009} that was applied symmetrically (radially) outward from the center of the CVD-grown graphene sheets prior to the thermal equilibration and subsequent velocity-driven actuation.  

\begin{figure} \begin{center} 
\includegraphics[scale=0.35]{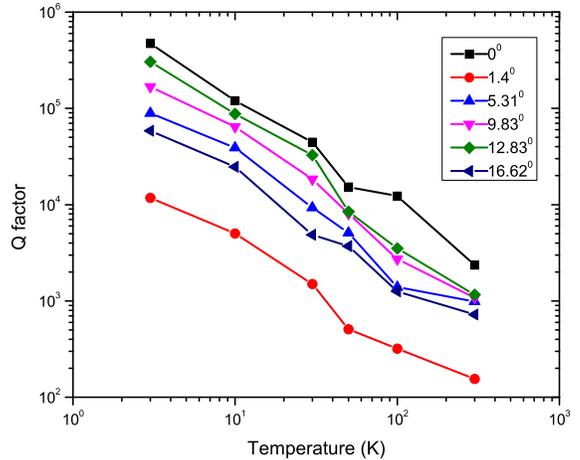}
\caption{\label{}(Color online)  Variation in the Q-factor for monolayer graphene under 1\% tensile strain as a function of temperature and grain boundary misorientation angle.}
\label{qstrain} \end{center} \end{figure}

\begin{figure} \begin{center} 
\includegraphics[scale=0.3]{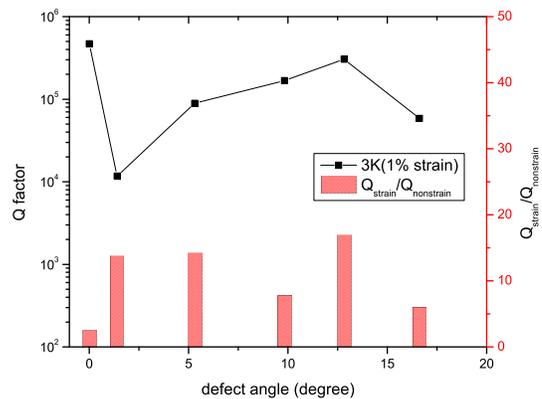}
\caption{\label{}(Color online)  Variation in the Q-factor as a function of grain boundary misorientation angle at 3K due to 1\% tensile strain, and the the ratio the Q-factors under tensile strain to the Q-factors of unstrained graphene.}
\label{qstrain3K} \end{center} \end{figure}

As shown by comparison of Figs. \ref{nostrainq} and \ref{qstrain}, the Q-factors of strained graphene with grain boundaries can, in some cases, approach those of pristine graphene.  In addition, as shown in Fig. \ref{qstrain3K}, tensile strain increases the Q-factors of graphene with grain boundaries much more than pristine graphene, where the tensile strain-induced Q-factor enhancement for the graphene with grain boundaries can be larger than one order of magnitude.  Fig. \ref{qstrain} also shows that strained graphene follows the same $Q \sim T^{-\alpha}$ relationship as unstrained graphene, though with a slightly lower exponent $\alpha$ of 1.11.  

We have found through our MD simulations that tensile strain increases the Q-factor by both increasing the natural frequency $\omega$, while simultaneously suppressing the damping $\gamma$ in the expression $\zeta=\gamma/\omega$, where Q$=0.5/\zeta$.  We found in analyzing the Q-factors that the tensile strain increased the natural frequency $\omega$ about the same amount for both pristine graphene and graphene with grain boundaries.  However, the damping factor $\gamma$ showed a significantly greater reduction for graphene with grain boundaries as compared to pristine graphene.  This is because the tensile strain reduces the out of plane buckling that results due to the 5-7 defect pairs, thus further enhancing the structural integrity and suppressing the damping of the oscillating graphene monolayer.  The specific buckling heights at 3K due to 1\% tensile strain for the grain boundary misorientation angles of 1.4$^{\circ}$, 5.31$^{\circ}$, 9.83$^{\circ}$, 12.83$^{\circ}$ and 16.62$^{\circ}$:  2.5\AA, 1.8\AA, 1.4\AA, 1.1\AA and 1.7\AA, respectively; these are clearly smaller than when no tensile strain is applied.

We also note that when tensile strain is applied the misorientation angle-dependent Q-factor follows a similar trend in Fig. \ref{qstrain3K} as was previously observed for the unstrained graphene in Fig. \ref{nostrainq}.  For strained graphene at 3K, we find that graphene with a misorientation angle of 12.83$^{\circ}$ has the highest Q factor, while the graphene sheets with a 9.83$^{\circ}$ misorientation angle has a similarly high Q-factor.  Overall, this demonstrates the important fact that CVD-grown graphene can, under modest, experimentally accessible tensile strains, exhibit comparable performance for NEMS sensing applications as pristine graphene.  

\section{Conclusions}

We have utilized classical MD simulations to quantify the effects of grain boundaries in CVD-grown graphene on the Q-factors of graphene nanoresonators.  Graphene with grain boundaries exhibit Q-factors that are 1-2 orders of magnitude smaller than pristine graphene.  However, the Q-factors follow a non-monotonic dependence on the grain boundary misorientation angle due to the competing effects of increased 5-7 defect pair density on one hand, and the increased structural symmetry and reduction in out of plane buckling heights on the other hand.  Furthermore, for practical applications, the Q-factors of CVD-grown graphene can be enhanced by about one order of magnitude through the application of 1\% tensile strain, which results in Q-factors that approach those of pristine graphene.

\section{Acknowledgements}

ZQ acknowledges support from a Boston University Dean's Catalyst Award.  HSP acknowledges support of NSF grant CMMI-0856261.




\footnotesize{
\bibliography{biball} 

\begin{thebibliography}{34}
\expandafter\ifx\csname natexlab\endcsname\relax\def\natexlab#1{#1}\fi
\expandafter\ifx\csname bibnamefont\endcsname\relax
  \def\bibnamefont#1{#1}\fi
\expandafter\ifx\csname bibfnamefont\endcsname\relax
  \def\bibfnamefont#1{#1}\fi
\expandafter\ifx\csname citenamefont\endcsname\relax
  \def\citenamefont#1{#1}\fi
\expandafter\ifx\csname url\endcsname\relax
  \def\url#1{\texttt{#1}}\fi
\expandafter\ifx\csname urlprefix\endcsname\relax\def\urlprefix{URL }\fi
\providecommand{\bibinfo}[2]{#2}
\providecommand{\eprint}[2][]{\url{#2}}

\bibitem[{\citenamefont{Novoselov et~al.}(2005)\citenamefont{Novoselov, Geim,
  Morozov, Jiang, Katsnelson, Grigorieva, Dubonos, and
  Firsov}}]{novoselovNATURE2005}
\bibinfo{author}{\bibfnamefont{K.~S.} \bibnamefont{Novoselov}},
  \bibinfo{author}{\bibfnamefont{A.~K.} \bibnamefont{Geim}},
  \bibinfo{author}{\bibfnamefont{S.~V.} \bibnamefont{Morozov}},
  \bibinfo{author}{\bibfnamefont{D.}~\bibnamefont{Jiang}},
  \bibinfo{author}{\bibfnamefont{M.~I.} \bibnamefont{Katsnelson}},
  \bibinfo{author}{\bibfnamefont{I.~V.} \bibnamefont{Grigorieva}},
  \bibinfo{author}{\bibfnamefont{S.~V.} \bibnamefont{Dubonos}},
  \bibnamefont{and} \bibinfo{author}{\bibfnamefont{A.~A.}
  \bibnamefont{Firsov}}, \bibinfo{journal}{Nature}
  \textbf{\bibinfo{volume}{438}}, \bibinfo{pages}{197} (\bibinfo{year}{2005}).

\bibitem[{\citenamefont{Geim and Novoselov}(2007)}]{geimNM2007}
\bibinfo{author}{\bibfnamefont{A.~K.} \bibnamefont{Geim}} \bibnamefont{and}
  \bibinfo{author}{\bibfnamefont{K.~S.} \bibnamefont{Novoselov}},
  \bibinfo{journal}{Nature Materials} \textbf{\bibinfo{volume}{6}},
  \bibinfo{pages}{183} (\bibinfo{year}{2007}).

\bibitem[{\citenamefont{Neto et~al.}(2009)\citenamefont{Neto, Guinea, Peres,
  Novoselov, and Geim}}]{netoRMP2009}
\bibinfo{author}{\bibfnamefont{A.~H.~C.} \bibnamefont{Neto}},
  \bibinfo{author}{\bibfnamefont{F.}~\bibnamefont{Guinea}},
  \bibinfo{author}{\bibfnamefont{N.~M.~R.} \bibnamefont{Peres}},
  \bibinfo{author}{\bibfnamefont{K.~S.} \bibnamefont{Novoselov}},
  \bibnamefont{and} \bibinfo{author}{\bibfnamefont{A.~K.} \bibnamefont{Geim}},
  \bibinfo{journal}{Reviews of Modern Physics} \textbf{\bibinfo{volume}{81}},
  \bibinfo{pages}{109} (\bibinfo{year}{2009}).

\bibitem[{\citenamefont{Han et~al.}(2007)\citenamefont{Han, Ozyilmaz, Zhang,
  and Kim}}]{hanPRL2007}
\bibinfo{author}{\bibfnamefont{M.~Y.} \bibnamefont{Han}},
  \bibinfo{author}{\bibfnamefont{B.}~\bibnamefont{Ozyilmaz}},
  \bibinfo{author}{\bibfnamefont{Y.}~\bibnamefont{Zhang}}, \bibnamefont{and}
  \bibinfo{author}{\bibfnamefont{P.}~\bibnamefont{Kim}},
  \bibinfo{journal}{Physical Review Letters} \textbf{\bibinfo{volume}{98}},
  \bibinfo{pages}{206805} (\bibinfo{year}{2007}).

\bibitem[{\citenamefont{Seol et~al.}(2010)\citenamefont{Seol, Jo, Moore,
  Lindsay, Aitken, Pettes, Li, Yao, Huang, Broido et~al.}}]{seolSCIENCE2010}
\bibinfo{author}{\bibfnamefont{J.~H.} \bibnamefont{Seol}},
  \bibinfo{author}{\bibfnamefont{I.}~\bibnamefont{Jo}},
  \bibinfo{author}{\bibfnamefont{A.~L.} \bibnamefont{Moore}},
  \bibinfo{author}{\bibfnamefont{L.}~\bibnamefont{Lindsay}},
  \bibinfo{author}{\bibfnamefont{Z.~H.} \bibnamefont{Aitken}},
  \bibinfo{author}{\bibfnamefont{M.~T.} \bibnamefont{Pettes}},
  \bibinfo{author}{\bibfnamefont{X.}~\bibnamefont{Li}},
  \bibinfo{author}{\bibfnamefont{Z.}~\bibnamefont{Yao}},
  \bibinfo{author}{\bibfnamefont{R.}~\bibnamefont{Huang}},
  \bibinfo{author}{\bibfnamefont{D.}~\bibnamefont{Broido}},
  \bibnamefont{et~al.}, \bibinfo{journal}{Science}
  \textbf{\bibinfo{volume}{328}}, \bibinfo{pages}{213} (\bibinfo{year}{2010}).

\bibitem[{\citenamefont{Bunch et~al.}(2007)\citenamefont{Bunch, van~der Zande,
  Verbridge, Frank, Tanenbaum, Parpia, Craighead, and
  McEuen}}]{bunchSCIENCE2007}
\bibinfo{author}{\bibfnamefont{J.~S.} \bibnamefont{Bunch}},
  \bibinfo{author}{\bibfnamefont{A.~M.} \bibnamefont{van~der Zande}},
  \bibinfo{author}{\bibfnamefont{S.~S.} \bibnamefont{Verbridge}},
  \bibinfo{author}{\bibfnamefont{I.~W.} \bibnamefont{Frank}},
  \bibinfo{author}{\bibfnamefont{D.~M.} \bibnamefont{Tanenbaum}},
  \bibinfo{author}{\bibfnamefont{J.~M.} \bibnamefont{Parpia}},
  \bibinfo{author}{\bibfnamefont{H.~G.} \bibnamefont{Craighead}},
  \bibnamefont{and} \bibinfo{author}{\bibfnamefont{P.~L.}
  \bibnamefont{McEuen}}, \bibinfo{journal}{Science}
  \textbf{\bibinfo{volume}{315}}, \bibinfo{pages}{490} (\bibinfo{year}{2007}).

\bibitem[{\citenamefont{Bunch et~al.}(2008)\citenamefont{Bunch, Verbridge,
  Alden, Zande, Parpia, Craighead, and McEuen}}]{bunchNL2008}
\bibinfo{author}{\bibfnamefont{J.~S.} \bibnamefont{Bunch}},
  \bibinfo{author}{\bibfnamefont{S.~S.} \bibnamefont{Verbridge}},
  \bibinfo{author}{\bibfnamefont{J.~S.} \bibnamefont{Alden}},
  \bibinfo{author}{\bibfnamefont{A.~M. V.~D.} \bibnamefont{Zande}},
  \bibinfo{author}{\bibfnamefont{J.~M.} \bibnamefont{Parpia}},
  \bibinfo{author}{\bibfnamefont{H.~G.} \bibnamefont{Craighead}},
  \bibnamefont{and} \bibinfo{author}{\bibfnamefont{P.~L.}
  \bibnamefont{McEuen}}, \bibinfo{journal}{Nano Letters}
  \textbf{\bibinfo{volume}{8}}, \bibinfo{pages}{2458} (\bibinfo{year}{2008}).

\bibitem[{\citenamefont{Robinson et~al.}(2008)\citenamefont{Robinson,
  Zalalutdinov, Baldwin, Snow, Wei, Sheehan, and Houston}}]{robinsonNL2008a}
\bibinfo{author}{\bibfnamefont{J.~T.} \bibnamefont{Robinson}},
  \bibinfo{author}{\bibfnamefont{M.}~\bibnamefont{Zalalutdinov}},
  \bibinfo{author}{\bibfnamefont{J.~W.} \bibnamefont{Baldwin}},
  \bibinfo{author}{\bibfnamefont{E.~S.} \bibnamefont{Snow}},
  \bibinfo{author}{\bibfnamefont{Z.}~\bibnamefont{Wei}},
  \bibinfo{author}{\bibfnamefont{P.}~\bibnamefont{Sheehan}}, \bibnamefont{and}
  \bibinfo{author}{\bibfnamefont{B.~H.} \bibnamefont{Houston}},
  \bibinfo{journal}{Nano Letters} \textbf{\bibinfo{volume}{8}},
  \bibinfo{pages}{3441} (\bibinfo{year}{2008}).

\bibitem[{\citenamefont{Barton et~al.}(2011{\natexlab{a}})\citenamefont{Barton,
  Parpia, and Craighead}}]{bartonJVSTB2011}
\bibinfo{author}{\bibfnamefont{R.~A.} \bibnamefont{Barton}},
  \bibinfo{author}{\bibfnamefont{J.}~\bibnamefont{Parpia}}, \bibnamefont{and}
  \bibinfo{author}{\bibfnamefont{H.~G.} \bibnamefont{Craighead}},
  \bibinfo{journal}{Journal of Vacuum Science and Technology B}
  \textbf{\bibinfo{volume}{29}}, \bibinfo{pages}{050801}
  (\bibinfo{year}{2011}{\natexlab{a}}).

\bibitem[{\citenamefont{Eom et~al.}(2011)\citenamefont{Eom, Park, Yoon, and
  Kwon}}]{eomPR2011}
\bibinfo{author}{\bibfnamefont{K.}~\bibnamefont{Eom}},
  \bibinfo{author}{\bibfnamefont{H.~S.} \bibnamefont{Park}},
  \bibinfo{author}{\bibfnamefont{D.~S.} \bibnamefont{Yoon}}, \bibnamefont{and}
  \bibinfo{author}{\bibfnamefont{T.}~\bibnamefont{Kwon}},
  \bibinfo{journal}{Physics Reports} \textbf{\bibinfo{volume}{503}},
  \bibinfo{pages}{115} (\bibinfo{year}{2011}).

\bibitem[{\citenamefont{Lee et~al.}(2008)\citenamefont{Lee, Wei, Kysar, and
  Hone}}]{leeSCIENCE2008}
\bibinfo{author}{\bibfnamefont{C.}~\bibnamefont{Lee}},
  \bibinfo{author}{\bibfnamefont{X.}~\bibnamefont{Wei}},
  \bibinfo{author}{\bibfnamefont{J.~W.} \bibnamefont{Kysar}}, \bibnamefont{and}
  \bibinfo{author}{\bibfnamefont{J.}~\bibnamefont{Hone}},
  \bibinfo{journal}{Science} \textbf{\bibinfo{volume}{321}},
  \bibinfo{pages}{385} (\bibinfo{year}{2008}).

\bibitem[{\citenamefont{Garcia-Sanchez
  et~al.}(2008)\citenamefont{Garcia-Sanchez, van~der Zande, Paulo, Lassagne,
  McEuen, and Bachtold}}]{sanchezNL2008}
\bibinfo{author}{\bibfnamefont{D.}~\bibnamefont{Garcia-Sanchez}},
  \bibinfo{author}{\bibfnamefont{A.~M.} \bibnamefont{van~der Zande}},
  \bibinfo{author}{\bibfnamefont{A.~S.} \bibnamefont{Paulo}},
  \bibinfo{author}{\bibfnamefont{B.}~\bibnamefont{Lassagne}},
  \bibinfo{author}{\bibfnamefont{P.~L.} \bibnamefont{McEuen}},
  \bibnamefont{and} \bibinfo{author}{\bibfnamefont{A.}~\bibnamefont{Bachtold}},
  \bibinfo{journal}{Nano Letters} \textbf{\bibinfo{volume}{8}},
  \bibinfo{pages}{1399} (\bibinfo{year}{2008}).

\bibitem[{\citenamefont{Kim and Park}(2009{\natexlab{a}})}]{kimNL2009}
\bibinfo{author}{\bibfnamefont{S.~Y.} \bibnamefont{Kim}} \bibnamefont{and}
  \bibinfo{author}{\bibfnamefont{H.~S.} \bibnamefont{Park}},
  \bibinfo{journal}{Nano Letters} \textbf{\bibinfo{volume}{9}},
  \bibinfo{pages}{969} (\bibinfo{year}{2009}{\natexlab{a}}).

\bibitem[{\citenamefont{Kim and Park}(2009{\natexlab{b}})}]{kimAPL2009}
\bibinfo{author}{\bibfnamefont{S.~Y.} \bibnamefont{Kim}} \bibnamefont{and}
  \bibinfo{author}{\bibfnamefont{H.~S.} \bibnamefont{Park}},
  \bibinfo{journal}{Applied Physics Letters} \textbf{\bibinfo{volume}{94}},
  \bibinfo{pages}{101918} (\bibinfo{year}{2009}{\natexlab{b}}).

\bibitem[{\citenamefont{Barton et~al.}(2011{\natexlab{b}})\citenamefont{Barton,
  Ilic, van~der Zande, Whitney, McEuen, Parpia, and Craighead}}]{bartonNL2011}
\bibinfo{author}{\bibfnamefont{R.~A.} \bibnamefont{Barton}},
  \bibinfo{author}{\bibfnamefont{B.}~\bibnamefont{Ilic}},
  \bibinfo{author}{\bibfnamefont{A.~M.} \bibnamefont{van~der Zande}},
  \bibinfo{author}{\bibfnamefont{W.~S.} \bibnamefont{Whitney}},
  \bibinfo{author}{\bibfnamefont{P.~L.} \bibnamefont{McEuen}},
  \bibinfo{author}{\bibfnamefont{J.~M.} \bibnamefont{Parpia}},
  \bibnamefont{and} \bibinfo{author}{\bibfnamefont{H.~G.}
  \bibnamefont{Craighead}}, \bibinfo{journal}{Nano Letters}
  \textbf{\bibinfo{volume}{11}}, \bibinfo{pages}{1232}
  (\bibinfo{year}{2011}{\natexlab{b}}).

\bibitem[{\citenamefont{Li et~al.}(2009)\citenamefont{Li, Cai, An, Kim, Nah,
  Yang, Piner, Velamakanni, Jung, Tutuc et~al.}}]{liSCIENCE2009}
\bibinfo{author}{\bibfnamefont{X.}~\bibnamefont{Li}},
  \bibinfo{author}{\bibfnamefont{W.}~\bibnamefont{Cai}},
  \bibinfo{author}{\bibfnamefont{J.}~\bibnamefont{An}},
  \bibinfo{author}{\bibfnamefont{S.}~\bibnamefont{Kim}},
  \bibinfo{author}{\bibfnamefont{J.}~\bibnamefont{Nah}},
  \bibinfo{author}{\bibfnamefont{D.}~\bibnamefont{Yang}},
  \bibinfo{author}{\bibfnamefont{R.}~\bibnamefont{Piner}},
  \bibinfo{author}{\bibfnamefont{A.}~\bibnamefont{Velamakanni}},
  \bibinfo{author}{\bibfnamefont{I.}~\bibnamefont{Jung}},
  \bibinfo{author}{\bibfnamefont{E.}~\bibnamefont{Tutuc}},
  \bibnamefont{et~al.}, \bibinfo{journal}{Science}
  \textbf{\bibinfo{volume}{324}}, \bibinfo{pages}{1312} (\bibinfo{year}{2009}).

\bibitem[{\citenamefont{Park and Ruoff}(2009)}]{parkNN2009}
\bibinfo{author}{\bibfnamefont{S.}~\bibnamefont{Park}} \bibnamefont{and}
  \bibinfo{author}{\bibfnamefont{R.~S.} \bibnamefont{Ruoff}},
  \bibinfo{journal}{Nature Nanotechnology} \textbf{\bibinfo{volume}{4}},
  \bibinfo{pages}{217} (\bibinfo{year}{2009}).

\bibitem[{\citenamefont{Reina et~al.}(2009)\citenamefont{Reina, Jia, Ho,
  Nezich, Son, Bulovic, Dresselhaus, and Kong}}]{reinaNL2009}
\bibinfo{author}{\bibfnamefont{A.}~\bibnamefont{Reina}},
  \bibinfo{author}{\bibfnamefont{X.}~\bibnamefont{Jia}},
  \bibinfo{author}{\bibfnamefont{J.}~\bibnamefont{Ho}},
  \bibinfo{author}{\bibfnamefont{D.}~\bibnamefont{Nezich}},
  \bibinfo{author}{\bibfnamefont{H.}~\bibnamefont{Son}},
  \bibinfo{author}{\bibfnamefont{V.}~\bibnamefont{Bulovic}},
  \bibinfo{author}{\bibfnamefont{M.~S.} \bibnamefont{Dresselhaus}},
  \bibnamefont{and} \bibinfo{author}{\bibfnamefont{J.}~\bibnamefont{Kong}},
  \bibinfo{journal}{Nano Letters} \textbf{\bibinfo{volume}{9}},
  \bibinfo{pages}{30} (\bibinfo{year}{2009}).

\bibitem[{\citenamefont{van~der Zande et~al.}(2010)\citenamefont{van~der Zande,
  Barton, Alden, Ruiz-Vargas, Whitney, Pham, Park, Parpia, Craighead, and
  McEuen}}]{zandeNL2010}
\bibinfo{author}{\bibfnamefont{A.~M.} \bibnamefont{van~der Zande}},
  \bibinfo{author}{\bibfnamefont{R.~A.} \bibnamefont{Barton}},
  \bibinfo{author}{\bibfnamefont{J.~S.} \bibnamefont{Alden}},
  \bibinfo{author}{\bibfnamefont{C.~S.} \bibnamefont{Ruiz-Vargas}},
  \bibinfo{author}{\bibfnamefont{W.~S.} \bibnamefont{Whitney}},
  \bibinfo{author}{\bibfnamefont{P.~H.~Q.} \bibnamefont{Pham}},
  \bibinfo{author}{\bibfnamefont{J.}~\bibnamefont{Park}},
  \bibinfo{author}{\bibfnamefont{J.~M.} \bibnamefont{Parpia}},
  \bibinfo{author}{\bibfnamefont{H.~G.} \bibnamefont{Craighead}},
  \bibnamefont{and} \bibinfo{author}{\bibfnamefont{P.~L.}
  \bibnamefont{McEuen}}, \bibinfo{journal}{Nano Letters}
  \textbf{\bibinfo{volume}{10}}, \bibinfo{pages}{4869} (\bibinfo{year}{2010}).

\bibitem[{\citenamefont{Kim et~al.}(2009)\citenamefont{Kim, Zhao, Jang, Lee,
  Kim, Kim, Ahn, Kim, Choi, and Hong}}]{kimNATURE2009}
\bibinfo{author}{\bibfnamefont{K.~S.} \bibnamefont{Kim}},
  \bibinfo{author}{\bibfnamefont{Y.}~\bibnamefont{Zhao}},
  \bibinfo{author}{\bibfnamefont{H.}~\bibnamefont{Jang}},
  \bibinfo{author}{\bibfnamefont{S.~Y.} \bibnamefont{Lee}},
  \bibinfo{author}{\bibfnamefont{J.~M.} \bibnamefont{Kim}},
  \bibinfo{author}{\bibfnamefont{K.~S.} \bibnamefont{Kim}},
  \bibinfo{author}{\bibfnamefont{J.-H.} \bibnamefont{Ahn}},
  \bibinfo{author}{\bibfnamefont{P.}~\bibnamefont{Kim}},
  \bibinfo{author}{\bibfnamefont{J.-Y.} \bibnamefont{Choi}}, \bibnamefont{and}
  \bibinfo{author}{\bibfnamefont{B.~H.} \bibnamefont{Hong}},
  \bibinfo{journal}{Nature} \textbf{\bibinfo{volume}{457}},
  \bibinfo{pages}{706} (\bibinfo{year}{2009}).

\bibitem[{\citenamefont{Huang et~al.}(2011)\citenamefont{Huang, Ruiz-Vargas,
  van~der Zande, Whitney, Levendorf, Kevek, Garg, Alden, Hustedt, Zhu
  et~al.}}]{huangNATURE2011}
\bibinfo{author}{\bibfnamefont{P.~Y.} \bibnamefont{Huang}},
  \bibinfo{author}{\bibfnamefont{C.~S.} \bibnamefont{Ruiz-Vargas}},
  \bibinfo{author}{\bibfnamefont{A.~M.} \bibnamefont{van~der Zande}},
  \bibinfo{author}{\bibfnamefont{W.~S.} \bibnamefont{Whitney}},
  \bibinfo{author}{\bibfnamefont{M.~P.} \bibnamefont{Levendorf}},
  \bibinfo{author}{\bibfnamefont{J.~W.} \bibnamefont{Kevek}},
  \bibinfo{author}{\bibfnamefont{S.}~\bibnamefont{Garg}},
  \bibinfo{author}{\bibfnamefont{J.~S.} \bibnamefont{Alden}},
  \bibinfo{author}{\bibfnamefont{C.~J.} \bibnamefont{Hustedt}},
  \bibinfo{author}{\bibfnamefont{Y.}~\bibnamefont{Zhu}}, \bibnamefont{et~al.},
  \bibinfo{journal}{Nature} \textbf{\bibinfo{volume}{469}},
  \bibinfo{pages}{389} (\bibinfo{year}{2011}).

\bibitem[{\citenamefont{Nemes-Incze et~al.}(2011)\citenamefont{Nemes-Incze,
  Yoo, Tapaszto, Dobrik, Labar, Horvath, Hwang, and Biro}}]{Nemes-Incze2011}
\bibinfo{author}{\bibfnamefont{P.}~\bibnamefont{Nemes-Incze}},
  \bibinfo{author}{\bibfnamefont{K.~J.} \bibnamefont{Yoo}},
  \bibinfo{author}{\bibfnamefont{L.}~\bibnamefont{Tapaszto}},
  \bibinfo{author}{\bibfnamefont{G.}~\bibnamefont{Dobrik}},
  \bibinfo{author}{\bibfnamefont{J.}~\bibnamefont{Labar}},
  \bibinfo{author}{\bibfnamefont{Z.~E.} \bibnamefont{Horvath}},
  \bibinfo{author}{\bibfnamefont{C.}~\bibnamefont{Hwang}}, \bibnamefont{and}
  \bibinfo{author}{\bibfnamefont{L.~P.} \bibnamefont{Biro}},
  \bibinfo{journal}{Applied Physics Letters} \textbf{\bibinfo{volume}{99}},
  \bibinfo{pages}{3} (\bibinfo{year}{2011}).

\bibitem[{\citenamefont{Carr and Lusk}(2010)}]{Carr2010}
\bibinfo{author}{\bibfnamefont{L.~D.} \bibnamefont{Carr}} \bibnamefont{and}
  \bibinfo{author}{\bibfnamefont{M.~T.} \bibnamefont{Lusk}},
  \bibinfo{journal}{Nature Nanotechnology} \textbf{\bibinfo{volume}{5}},
  \bibinfo{pages}{316} (\bibinfo{year}{2010}).

\bibitem[{\citenamefont{Kim and Park}(2010)}]{kimNANO2010}
\bibinfo{author}{\bibfnamefont{S.~Y.} \bibnamefont{Kim}} \bibnamefont{and}
  \bibinfo{author}{\bibfnamefont{H.~S.} \bibnamefont{Park}},
  \bibinfo{journal}{Nanotechnology} \textbf{\bibinfo{volume}{21}},
  \bibinfo{pages}{105710} (\bibinfo{year}{2010}).

\bibitem[{\citenamefont{Stuart et~al.}(2000)\citenamefont{Stuart, Tutein, and
  Harrison}}]{stuartJCP2000}
\bibinfo{author}{\bibfnamefont{S.~J.} \bibnamefont{Stuart}},
  \bibinfo{author}{\bibfnamefont{A.~B.} \bibnamefont{Tutein}},
  \bibnamefont{and} \bibinfo{author}{\bibfnamefont{J.~A.}
  \bibnamefont{Harrison}}, \bibinfo{journal}{Journal of Chemical Physics}
  \textbf{\bibinfo{volume}{112}}, \bibinfo{pages}{6472} (\bibinfo{year}{2000}).

\bibitem[{\citenamefont{Lammps}(2012)}]{plimptonLAMMPS}
\bibinfo{author}{\bibnamefont{Lammps}},
  \bibinfo{journal}{http://lammps.sandia.gov}  (\bibinfo{year}{2012}).

\bibitem[{\citenamefont{Hoover}(1985)}]{hooverPRA1985}
\bibinfo{author}{\bibfnamefont{W.~G.} \bibnamefont{Hoover}},
  \bibinfo{journal}{Physical Review A} \textbf{\bibinfo{volume}{31}},
  \bibinfo{pages}{1695} (\bibinfo{year}{1985}).

\bibitem[{\citenamefont{Liu et~al.}(2011)\citenamefont{Liu, Gajewski, Pao, and
  Chang}}]{Liu2011}
\bibinfo{author}{\bibfnamefont{T.~H.} \bibnamefont{Liu}},
  \bibinfo{author}{\bibfnamefont{G.}~\bibnamefont{Gajewski}},
  \bibinfo{author}{\bibfnamefont{C.~W.} \bibnamefont{Pao}}, \bibnamefont{and}
  \bibinfo{author}{\bibfnamefont{C.~C.} \bibnamefont{Chang}},
  \bibinfo{journal}{Carbon} \textbf{\bibinfo{volume}{49}},
  \bibinfo{pages}{2306} (\bibinfo{year}{2011}).

\bibitem[{\citenamefont{Humphrey et~al.}(1996)\citenamefont{Humphrey, Dalke,
  and Schulten}}]{Humphrey1996}
\bibinfo{author}{\bibfnamefont{W.}~\bibnamefont{Humphrey}},
  \bibinfo{author}{\bibfnamefont{A.}~\bibnamefont{Dalke}}, \bibnamefont{and}
  \bibinfo{author}{\bibfnamefont{K.}~\bibnamefont{Schulten}},
  \bibinfo{journal}{Journal of Molecular Graphics}
  \textbf{\bibinfo{volume}{14}}, \bibinfo{pages}{33} (\bibinfo{year}{1996}).

\bibitem[{\citenamefont{Vallabhaneni et~al.}(2011)\citenamefont{Vallabhaneni,
  Rhoads, Murthy, and Ruan}}]{valJAP2011}
\bibinfo{author}{\bibfnamefont{A.~K.} \bibnamefont{Vallabhaneni}},
  \bibinfo{author}{\bibfnamefont{J.~F.} \bibnamefont{Rhoads}},
  \bibinfo{author}{\bibfnamefont{J.~Y.} \bibnamefont{Murthy}},
  \bibnamefont{and} \bibinfo{author}{\bibfnamefont{X.}~\bibnamefont{Ruan}},
  \bibinfo{journal}{Journal of Applied Physics} \textbf{\bibinfo{volume}{110}},
  \bibinfo{pages}{034312} (\bibinfo{year}{2011}).

\bibitem[{\citenamefont{Chu et~al.}(2007)\citenamefont{Chu, Rudd, and
  Blencowe}}]{chuARXIV2007}
\bibinfo{author}{\bibfnamefont{M.}~\bibnamefont{Chu}},
  \bibinfo{author}{\bibfnamefont{R.~E.} \bibnamefont{Rudd}}, \bibnamefont{and}
  \bibinfo{author}{\bibfnamefont{M.~P.} \bibnamefont{Blencowe}},
  \bibinfo{journal}{Arxiv.org} p. \bibinfo{pages}{0705.0015v1}
  (\bibinfo{year}{2007}).

\bibitem[{\citenamefont{Verbridge et~al.}(2007)\citenamefont{Verbridge,
  Shapiro, Craighead, and Parpia}}]{verbridgeNL2007}
\bibinfo{author}{\bibfnamefont{S.~S.} \bibnamefont{Verbridge}},
  \bibinfo{author}{\bibfnamefont{D.~F.} \bibnamefont{Shapiro}},
  \bibinfo{author}{\bibfnamefont{H.~G.} \bibnamefont{Craighead}},
  \bibnamefont{and} \bibinfo{author}{\bibfnamefont{J.~M.}
  \bibnamefont{Parpia}}, \bibinfo{journal}{Nano Letters}
  \textbf{\bibinfo{volume}{7}}, \bibinfo{pages}{1728} (\bibinfo{year}{2007}).

\bibitem[{\citenamefont{Huang et~al.}(2009)\citenamefont{Huang, Yan, Chen,
  Song, Heinz, and Hone}}]{Huang2009}
\bibinfo{author}{\bibfnamefont{M.}~\bibnamefont{Huang}},
  \bibinfo{author}{\bibfnamefont{H.}~\bibnamefont{Yan}},
  \bibinfo{author}{\bibfnamefont{C.}~\bibnamefont{Chen}},
  \bibinfo{author}{\bibfnamefont{D.}~\bibnamefont{Song}},
  \bibinfo{author}{\bibfnamefont{T.~F.} \bibnamefont{Heinz}}, \bibnamefont{and}
  \bibinfo{author}{\bibfnamefont{J.}~\bibnamefont{Hone}},
  \bibinfo{journal}{Proceedings of the National Academy of Sciences of the
  United States of America} \textbf{\bibinfo{volume}{106}},
  \bibinfo{pages}{7304} (\bibinfo{year}{2009}).

\bibitem[{\citenamefont{Mohiuddin et~al.}(2009)\citenamefont{Mohiuddin,
  Lombardo, Nair, Bonetti, Savini, Jalil, Bonini, Basko, Galiotis, Marzari
  et~al.}}]{Mohiuddin2009}
\bibinfo{author}{\bibfnamefont{T.~M.~G.} \bibnamefont{Mohiuddin}},
  \bibinfo{author}{\bibfnamefont{A.}~\bibnamefont{Lombardo}},
  \bibinfo{author}{\bibfnamefont{R.~R.} \bibnamefont{Nair}},
  \bibinfo{author}{\bibfnamefont{A.}~\bibnamefont{Bonetti}},
  \bibinfo{author}{\bibfnamefont{G.}~\bibnamefont{Savini}},
  \bibinfo{author}{\bibfnamefont{R.}~\bibnamefont{Jalil}},
  \bibinfo{author}{\bibfnamefont{N.}~\bibnamefont{Bonini}},
  \bibinfo{author}{\bibfnamefont{D.~M.} \bibnamefont{Basko}},
  \bibinfo{author}{\bibfnamefont{C.}~\bibnamefont{Galiotis}},
  \bibinfo{author}{\bibfnamefont{N.}~\bibnamefont{Marzari}},
  \bibnamefont{et~al.}, \bibinfo{journal}{Physical Review B}
  \textbf{\bibinfo{volume}{79}}, \bibinfo{pages}{205433}
  (\bibinfo{year}{2009}).

\end{thebibliography}
\bibliographystyle{rsc} 
}

\end{document}